\documentclass[aps,prl,twocolumn,groupedaddress,superscriptaddress,nofootinbib,notitlepage,floatfix,showpacs,amssymb,amsmath,amsfonts]{revtex4-2}

\usepackage{times,bm,bbm,bbold,amssymb,amsmath,amsfonts,dsfont,graphics,graphicx,braket,color,xcolor,hyperref}

\hypersetup{colorlinks,linkcolor={blue},citecolor={blue},urlcolor= {blue}}
\DeclareFontFamily{OT1}{pzc}{}
\DeclareFontShape{OT1}{pzc}{m}{it}
{<-> s * [1.25] pzcmi7t}{}
\DeclareMathAlphabet{\mathpzc}{OT1}{pzc}
{m}{it}

\newcommand{\ignore}[1]{}
\usepackage[T1]{fontenc}
\usepackage{newtxtext,newtxmath}

\begin{document}
	

\title{Reducing the Number of Qubits from $n^2$ to $n\log_{2} (n)$ to Solve the Traveling Salesman Problem with Quantum Computers: A Proposal for Demonstrating Quantum Supremacy in the NISQ Era}

\author{Mehdi Ramezani}
\affiliation{Department of Physics, Sharif University of Technology, Tehran 14588, Iran}
\affiliation{Centre for Quantum Engineering and Photonics Technology, Sharif University of Technology, Tehran 14588, Iran}

\author{Sadegh Salami}
\affiliation{Centre for Quantum Engineering and Photonics Technology, Sharif University of Technology, Tehran 14588, Iran}
\affiliation{Department of Computer Engineering, Sharif University of Technology, Tehran 14588, Iran}

\author{Mehdi Shokhmkar}
\affiliation{Department of Physics, Sharif University of Technology, Tehran 14588, Iran}
\affiliation{Centre for Quantum Engineering and Photonics Technology, Sharif University of Technology, Tehran 14588, Iran}

\author{Morteza Moradi}
\affiliation{Institute of Informatics, Faculty of Mathematics, Informatics and Mechanics, University of Warsaw, Banacha 2, 02-097 Warsaw, Poland}

\author{Alireza Bahrampour}
\affiliation{Department of Physics, Sharif University of Technology, Tehran 14588, Iran}
\affiliation{Centre for Quantum Engineering and Photonics Technology, Sharif University of Technology, Tehran 14588, Iran}

\date{\today}

\begin{abstract}
	\noindent In our pursuit of quantum supremacy during the NISQ era, this research introduces a novel approach rooted in the Quantum Approximate Optimization Algorithm (QAOA) framework to address the Traveling Salesman Problem (TSP). By strategically reducing the requisite qubit count from $n^2$ to $n\log_{2} (n)$, our QAOA-based algorithm not only contributes to the ongoing discourse on qubit efficiency but also demonstrates improved performance based on established metrics, underscoring its potential for achieving NISQ-era supremacy in solving real-world optimization challenges.
\end{abstract}

\date{\today}
\maketitle


\textit{Introduction.---}The advent of Noisy Intermediate-Scale Quantum (NISQ) computing marks a pivotal phase in quantum technology, presenting both opportunities and challenges. As researchers strive to achieve quantum supremacy—where quantum computers surpass classical counterparts in specific tasks—explorations have primarily centered around sampling problems \cite{arute2019quantum,wu2021strong}. However, the practical implications of these achievements, particularly in domains like combinatorial optimization, remain a critical aspect to address. 

Combinatorial Optimization Problems (COPs) \cite{du1998handbook}, form a critical class of challenges where the goal is to find the best solution from a finite set of discrete possibilities. These problems have applications in various fields, including logistics, scheduling, and network design. COPs are notorious for their computational hardness \cite{ausiello2012complexity}. The difficulty stems from the exponential growth in the number of possible solutions as the size of the problem increases. As a result, the search space becomes vast, and finding the optimal solution becomes increasingly complex and time-consuming. 

The Traveling Salesman Problem (TSP) \cite{gavish1978travelling,dahiya2018literature} is a quintessential example within the realm of COPs. In the TSP, the objective is to determine the most efficient route that visits each of the given cities exactly once and returns to the starting city. This problem encapsulates the challenge of optimizing a permutation of cities, and as the number of cities increases, the number of possible routes grows factorially. Specifically, for the TSP with $n$ cities, the number of possible routes is given by $(n-1)!/2$, underscoring the combinatorial explosion inherent in such optimization challenges. 

Classical algorithms such as greedy algorithm \cite{kizilatecs2013nearest} and genetic algorithm \cite{potvin1996genetic,chatterjee1996genetic} often face a trade-off between the accuracy of the solution and the computational speed. Achieving high accuracy in finding optimal solutions typically involves intensive computation, limiting the applicability of these methods to real-time decision-making scenarios \cite{halim2019combinatorial}.

The best-known exact algorithm for solving the TSP is the Held-Karp algorithm \cite{held1962dynamic}, also known as dynamic programming approach \cite{bellman1962dynamic}. This algorithm computes the shortest tour that visits each city exactly once by considering all possible subsets of cities and storing intermediate results to avoid redundant calculations. The time complexity of the Held-Karp algorithm is $O(n^{2}2^{n})$.

Quantum computation offers promising avenues for solving COPs more efficiently, presenting a paradigm shift from classical approaches \cite{gemeinhardt2023quantum}. Among the various quantum algorithms, two main methods have emerged as particularly relevant for addressing COPs:

\textit{Variational Quantum Algorithms:} Notably, the Quantum Approximate Optimization Algorithm (QAOA) \cite{farhi2014quantum} falls under this category. Variational algorithms \cite{cerezo2021variational} transform the discrete solution space of COPs into a continuous parameter space of a quantum circuit. These algorithims, leverage a Hybrid method, wherein a classical computer explores the continuous parameter space while a quantum computer performs calculations to evaluate the cost function. This synergistic approach allows for more efficient exploration of potential solutions.

\textit{Quantum Annealing:} Another prominent method involves quantum annealing, a quantum computing approach that directly seeks solutions to problems without requiring a classical counterpart during the computation \cite{das2005quantum,hen2016quantum}. 

In both QAOA and quantum annealing, the sought-after solution corresponds to the ground state of the Hamiltonian, a fundamental concept in quantum mechanics. However, the methodologies differ; QAOA achieves this by navigating through the parameter space, while quantum annealing exploits the quantum adiabatic process to directly obtain the ground state of the Hamiltonian using suitble quantum hardware. This distinction underscores the versatility of quantum computing approaches in addressing COPs, offering diverse strategies to tackle combinatorial optimization challenges efficiently.

Indeed, while quantum annealing presents a compelling approach due to its inherent simplicity—bypassing the need to search through a parameterized space—it faces a notable challenge in the constraints imposed by current quantum hardware. Quantum computers, particularly those designed for quantum annealing, such as those developed by companies like D-Wave, currently exhibit limitations \cite{lenkiewicz2020d}. The hardware constraints include the ability to handle only \textit{quadratic} Hamiltonians, which essentially means that the Hamiltonian can be at most quadratic in the binary variables representing the problem.

Traditionally, for a TSP with $n$ cities, the standard formulation involves the introduction of $n^2$ Binary Variables (BVs), leading to a quadratic Hamiltonian in BVs \cite{lucas2014ising}. In this Letter, we redefine the TSP formulation to streamline the representation of solutions using a significantly reduced number of BVs—specifically, $n\log_{2} (n)$. This reduction in the number of BVs is a substantial advancement, demonstrating an efficient quantum algorithm tailored for contemporary quantum computers. However, it's noteworthy that our proposed Hamiltonian departs from the conventional quadratic structure associated with TSP formulations.

\textit{Quantum Approximate Optimization Algorithm.---}The QAOA stands as a versatile quantum paradigm for addressing combinatorial optimization problems within a hybrid quantum-classical framework. Introduced by Farhi, Goldstone, and Gutmann in 2014 \cite{farhi2014quantum}, QAOA excels in tackling intricate optimization challenges, including the TSP.

At the core of QAOA lies the parametrized quantum circuit $U(\bm{\gamma}, \bm{\beta})$, where $\bm{\gamma}$ and $\bm{\beta}$ are sets of variational parameters. This circuit is composed as a sequence of unitary operators representing the evolution of the quantum state under the influence of the mixing Hamiltonian $H_M$ and the problem Hamiltonian $H_P$. The parametrized quantum circuit takes the form:
\begin{equation}
	U(\bm{\gamma}, \bm{\beta}) = \prod_{l=1}^{L} \exp\left(-i\beta_{l}H_{M}\right)\exp\left(-i\gamma_{l}H_{P}\right),
\end{equation}
where $L$ denotes the number of optimization layers. The mixing Hamiltonian $H_M=\sum_{i=1}^{N} \sigma_{x}^{(i)}$, where $\sigma_{x}^{(i)}$ denotes the Pauli-X matrix acting on the $i-$th qubit, facilitates exploration across the solution space, while the problem Hamiltonian $H_P$ encapsulates the objective of the optimization problem.

At the beginning of the algorithm, the initial state of QAOA is the superposition of all bitstrings, which can be obtained by applying a Hadamard gate on each qubit. The QAOA seeks to optimize the parameters $\{\bm{\gamma}, \bm{\beta}\}$ to minimize the expectation value of the resulting quantum state with respect to the Hamiltonian $H_{P}$. This process leads to the identification of optimal solutions for the combinatorial optimization problem.

\textit{Standard BV formulation of TSP.}\textbf{---} The conventional BV formulation for the TSP is expressed as follows: Consider the binary variable $x_{i,j}$, which takes the value $1$ if city $i$ is visited at time $j$ and $0$ otherwise. The cost function (energy function) associated with the BVs is defined as $E_{P}=E_{C}+E_{R}$, where
\begin{equation}
	E_{C} = A \sum_{\nu=1}^{n}\left(1-\sum_{j=1}^{n} x_{\nu,j}\right)^{2} + A \sum_{j=1}^{n}\left(1-\sum_{\nu=1}^{n} x_{\nu,j}\right)^{2},
\end{equation}
represents the \textit{constraint energy}. The initial term ensures that each city is visited exactly once, while the subsequent term guarantees that at any given time, only one city is visited. Additionally,
\begin{equation}
	E_{R} = B \sum_{u,v} W_{u,v}\sum_{j=1}^{n}x_{u,j}x_{v,j+1},
\end{equation}
constitutes the \textit{route energy}, where $W_{u,v}$ denotes the distance from city $u$ to city $v$. This energy component captures the total distance of any given tour, providing a comprehensive measure of the optimization objective for the TSP. The coefficients $A$ and $B$ modulate the impact of the constraint and problem energies, respectively, influencing the overall landscape of the optimization function. 

To formulate a Hamiltonian corresponding to the aforementioned energy function, a straightforward substitution suffices. Simply replace any binary variable $x$ in the expression with the matrix $(\mathds{1}-\sigma_z)/2$, where $\sigma_z$ represents the Pauli-Z matrix, and $\mathds{1}$ is the identity matrix of size $2\times2$. This substitution ensures a seamless translation from the binary variable representation to the quantum framework, aligning the classical energy function with its quantum counterpart in the context of quantum annealing or variational quantum algorithms.

\textit{Our Method.---}In the conventional BV formulation of the Traveling TSP, a set of binary variables $\{x_{1,j}, x_{2,j}, \dots, x_{n,j}\}$ is utilized for each instance of time, denoted as $j$, to encode the identity of the city. The binary strings $\{1,0,0,\dots,0\}$, $\{0,1,0,\dots,0\}$, $\dots$, $\{0,0,0,\dots,1\}$ correspond to city numbers $1, 2, \dots, n$, respectively. However, the question arises: is dedicating $n$ qubits too extensive for encoding $n$ cities? It seems more judicious to leverage $\log_{2}(n)$ qubits to encode the binary representation of each city $\nu$, considering that $\log_{2}(n)$ bits are typically sufficient for encoding any integer from $1$ to $n$.

In their pioneering work \cite{vargas2021many}, a quadratic model for the TSP was introduced, departing from the conventional use of qubits by employing qudits. Remarkably, the study demonstrates that employing $n$ qudits with $d=n$ where the dimension of the Hilbert space is $n^n=2^{n\log_{2}(n)}$, allows the construction of a Hamiltonian. Notably, the ground state of this Hamiltonian corresponds to the optimal solution of the TSP. This novel approach supports the intriguing notion that utilizing $n\log_{2}(n)$ qubits may indeed suffice for formulating a BV of the TSP.

Consider $k$ as the number of qubits necessary to encode $n$ cities at each instance of time. In this context,, the BVs for the TSP are denoted as $\mathbf{x}=\{x_{i,j}\}$, where $i\in \{1,2,\dots,k\}$ and $j\in\{1,2,\dots,n\}$. Defining $\mathbf{x}_{i}=\{x_{1,i},x_{2,i},\dots,x_{k,i}\}$, our proposed energy function for the TSP is expressed as:
\begin{equation}
	E_{P}(\mathbf{x}) = E_C(\mathbf{x}) + E_R(\mathbf{x})
\end{equation}
where
\begin{align}
	E_C(\mathbf{x}) &= \sum_{i} E_{i,i}(\mathbf{x}_{i}) + \sum_{j>i+1} E_{i,j}(\mathbf{x}_{i},\mathbf{x}_{j}),\\
	E_R(\mathbf{x}) &= \sum_{i} E_{i,i+1}(\mathbf{x}_{i},\mathbf{x}_{i+1}).
\end{align}
The first terms of the \textit{constraint energy} $(E_{i,i}(\mathbf{x}_{i}))$ assign negative energy for $\mathbf{x}_{i}$ that corresponds to a valid encoding of cities at time $i$ and positive energy (penalty) for $\mathbf{x}_{i}$ that does not represent a valid city encoding. For $\mathbf{x}_{i}$ that corresponds to a valid city encoding, the second terms of the \textit{constraint energy} $(E_{i,j}(\mathbf{x}_{i},\mathbf{x}_{j}))$ introduce penalties when $\mathbf{x}_{i}=\mathbf{x}_{j}$ and yield zero when $\mathbf{x}_{i}\neq\mathbf{x}_{j}$. Lastly, each term of the \textit{route energy} $(E_{i,i+1}(\mathbf{x}_{i},\mathbf{x}_{i+1}))$ imposes penalties when $\mathbf{x}_{i}=\mathbf{x}_{i+1}$ and yields $W_{u,v}$ when $\mathbf{x}_{i}$ and $\mathbf{x}_{i+1}$ correspond to the encodings of cities $u$ and $v$ respectively. This energy function encapsulates the essential aspects of the TSP, providing a comprehensive representation of constraints and objectives within the quantum computational framework.

Define ${\mathbf{y}=\{y_m\}=\{\mathbf{x}_{i},\mathbf{x}_{j}\}}$ with ${m\in\{1,2,\dots,2k\}}$, in this case, the general form of $E_{i,j}(\mathbf{y})$  can be expressed as:
\begin{equation}\label{Eij}
	\begin{aligned}
		E_{i,j}(\mathbf{y}) &= a_{0} + \sum_{m_{1}} a_{m_{1}} y_{m_{1}} + \sum_{m_{2}>m_{1}} a_{m_{1},m_{2}} y_{m_{1}}y_{m_{2}}\\
		&+ \sum_{m_{3}>m_{2}>m_{1}} a_{m_{1},m_{2},m_{3}} y_{m_{1}}y_{m_{2}}y_{m_{3}}\\
		&+ \dots + a_{1,2,\dots,2k}y_{1}y_{2}\dots y_{2k}.
	\end{aligned}
\end{equation}
The number of parameters in this expression is equal to $\sum_{p=0}^{2k}\binom{2k}{p}=2^{2k}$. To satisfy the requirement of characterizing $E_{i,j}(\mathbf{y})$ with $n^{2}$ equations, one finds ${k=\log_{2}(n)}$. Hence, the required number of qubits would be ${N=nk=n\log_{2}(n)}$.

In the case where $k = \log_{2}(n)$, a natural encoding strategy for city numbers involves utilizing the binary representation of each city. For instance, in a TSP with 4 cities, the encoding $\{0,0\}, \{0,1\}, \{1,0\}, \{1,1\}$ would straightforwardly correspond to cities 1 to 4, respectively. In this scenario, every encoding represents a valid city number, obviating the necessity for the term $E_{i,i}(\mathbf{x}_{i})$ in the constraint energy. 

To succinctly outline our method, we construct the BV formulation of the TSP by introducing binary variables $\mathbf{x}=\{x_{i,j}\}$ where $i\in \{1,2,\dots,\log_{2}(n)\}$ and $j\in\{1,2,\dots,n\}$. The string $\mathbf{x}_{i}=\{x_{1,i},x_{2,i},\dots,x_{\log_{2}(n),i}\}$ corresponds to the encoding of cities at time step $i$. Notably, the encoding of city $\nu$ aligns with the binary representation of the number $\nu-1$. The energy function associated with the problem is given by:
\begin{equation}
	E_{P}(\mathbf{x}) = \sum_{j>i} E_{i,j}(\mathbf{x}_{i},\mathbf{x}_{j}),
\end{equation}
where $E_{i,j}(\mathbf{x}_{i},\mathbf{x}_{j})$ has the form of Eq.\eqref{Eij}, and to find the coefficients appear in this equation one should solve following system of linear equations for each $\mathbf{x}_{i}$ and $\mathbf{x}_{j}$:
\begin{equation}\label{Eij_equations}
	\begin{aligned}
		E_{i,j}(\mathbf{x}_{i},\mathbf{x}_{j})&=\delta_{i,j-1}\left( \delta_{\mathbf{x}_{i},\mathbf{x}_{j}}P + \left( 1- \delta_{\mathbf{x}_{i},\mathbf{x}_{j}} \right)W_{u,v} \right)\\
		&+ \left(1-\delta_{i,j-1}\right)\delta_{\mathbf{x}_{i},\mathbf{x}_{j}}P,
	\end{aligned}
\end{equation}
where $P$ represents the penalty term, and $\{u,v\}$  are the cities corresponding to the encoding $\{\mathbf{x}_{i},\mathbf{x}_{j}\}$ respectively. The solvability and uniqueness of the solution to this system of linear equations are discussed in the Supplementary Material section. This formulation captures the essential aspects of the TSP, providing a comprehensive representation of constraints and objectives within the quantum computational framework.

\textit{Simulation Results.---}In this section, we present the outcomes of our simulations comparing the performance of our $n\log_{2} (n)$ qubit encoding with the traditional $n^2$ qubit encoding. We assess the performance using standard evaluation metrics \cite{qian2023comparative}:

\noindent\textbf{Approximation Ratio (AR)}: This metric represents the ratio of the expectation value of the Hamiltonian $H_{P}$ in the output state of the algorithm to the distance of the best path, which is the optimal solution.

\noindent\textbf{True Percentage}: This metric represents the probability of the optimal solution, among the outcomes produced by the algorithm.

\noindent\textbf{Rank}: This metric indicates the position of the optimal solution relative to other states in terms of their probability.

To optimize the QAOA, we implemented the layerwise learning protocol as introduced in \cite{skolik2021layerwise}. In this approach, for a $p-$layer QAOA simulation, we fix the parameters in the first $(p-1)$ layers, which are obtained from previous simulations, and focus solely on optimizing the parameters in the $p-$th layer. To determine the optimal values for the parameters of each layer, we employed the Basin-Hopping algorithm \cite{wales1997global} with 500 iterations. The Basin-Hopping algorithm is particularly well-suited for finding the global minimum of an objective function.

\begin{figure*}[t]
	\includegraphics[width=\textwidth,height=5cm]{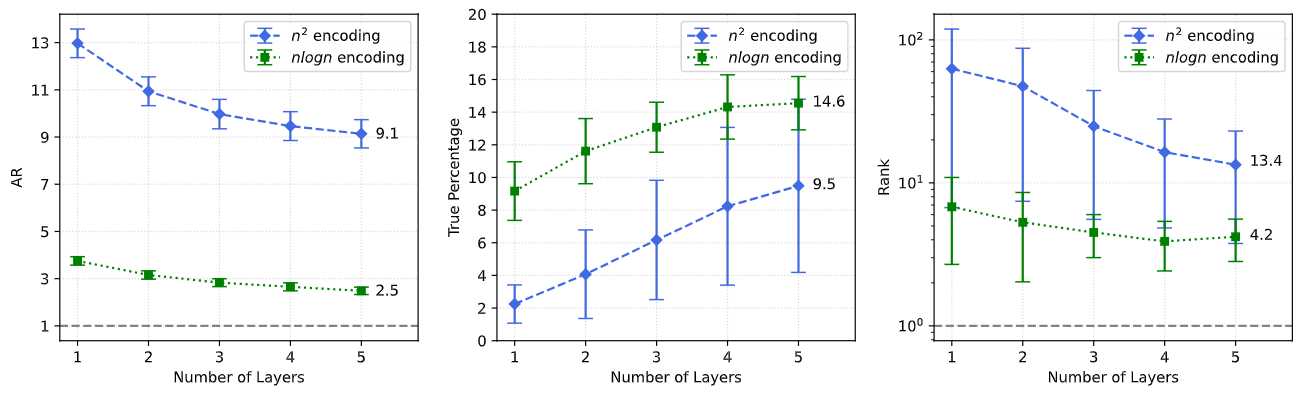}
	\caption{Comparison of evaluation metrics for $n \log_{2} (n)$ and $n^{2}$ encodings across different numbers of layers in QAOA simulation}
	\label{fig:results}
\end{figure*}

We conducted simulations of the QAOA algorithm for 10 samples of 4 cities. To create challenging instances of the TSP, we aimed to avoid clustering the cities within a specific region. To achieve this, we divided the $100\times100$ xy-plane into four $50\times50$ regions. For each city, we randomly selected its coordinates from each region using a normal distribution. The mean of the distribution was centered at the midpoint of the region, and the variance was set to 10. This approach ensured that the cities were distributed across the entire plane, presenting a more challenging TSP scenario for our simulations. 

Considering that the first city in the TSP is fixed, our encoding requires $(n-1)\log_{2} (n)$ qubits, while the conventional encoding demands $(n-1)^2$ qubits. Therefore, for a TSP instance with 4 cities, we utilized 6 qubits with our encoding and 9 qubits with the conventinal encoding.

The results of these simulations are illustrated in FIG. \ref{fig:results}. It is evident from the figure that the $n \log_{2} (n)$ encoding consistently outperforms the $n^{2}$ encoding across all evaluation metrics.

In this study, we compare the simulation results of $n \log_{2} (n)$ and $n^{2}$ encodings using the original version of the QAOA algorithm. However, for optimization problems with hard constraints such as the TSP involving a large number of cities, it is often more effective to utilize variant versions of the QAOA tailored specifically for TSP, as introduced in \cite{hadfield2017quantum,larose2022mixer}. These variant versions of QAOA designed for TSP incorporate different mixing Hamiltonians \cite{hadfield2019quantum} along with distinct initial states \cite{bartschi2020grover}. It is noteworthy that the proposed variants are primarily tailored for $n^{2}$ encoding. Exploring and proposing new variants specifically suited for $n \log_{2} (n)$ encoding could be a promising avenue for future research endeavors.

At the culmination of our analysis, it becomes imperative to contrast the gate complexity between the two encodings. While our proposed method effectively reduces the qubit count from $n^2$ to $n \log_{2} (n)$, it comes at the expense of higher gate complexity compared to the conventional encoding. Specifically, implementing the unitary gate of the problem Hamiltonian in $n^2$ encoding necessitates a scaling of $n^3$ two-qubit gates, whereas in our approach, this scales with $n^4 \log_{2} (n)$.

Moreover, two critical considerations warrant attention. Firstly, in practical quantum computers, the application of two-qubit gates is confined to neighboring qubits. Thus, for distant qubits, entanglement swapping techniques may be required for connectivity. Given our method's utilization of fewer qubits, it stands to reason that it would demand a lower count of two-qubit gates in real-world quantum computing applications, presenting a practical advantage for our proposed $n \log_{2} (n)$ encoding method.

Secondly, as discussed earlier, in tackling optimization problems like the TSP, leveraging variant versions of the QAOA proves more effective. In these variants, the limiting factor shifts from the unitary gate of the problem Hamiltonian to that of the mixing Hamiltonian. For instance, in the \textit{row swap} mixing Hamiltonian introduced in \cite{hadfield2017quantum}, the implementation of the unitary gate requires $O(m^2)$ four-qubit gates, where $m$ denotes the total number of qubits used in QAOA. This observation underscores the correlation between qubit reduction and the decrease in the number of multi-qubit gates, highlighting the potential efficiency gains in such cases.

The codes used for the simulations and the resulting data are publicly available on GitHub at the following address: \href{https://github.com/PsiAlgorithms/TSP_QAOA}{github.com/PsiAlgorithms/tsp\_qaoa}. Researchers and enthusiasts are encouraged to explore and utilize these resources for further study and experimentation in the field of quantum optimization.

\textit{Acknowledgements.---}This work was supported by the Research Centre for Quantum Engineering and Photonics Technology, Sharif University of Technology, through the Quantum NISQ Algorithms Project under Grant No. 140200402. M.M was supported by the European Union's Horizon 2020 research and innovation programme under the Marie Skłodowska-Curie project "AppQInfo" No. 956071.

\bibliographystyle{unsrt}
\bibliography{bibliography}

\onecolumngrid
\newpage
\appendix

\setcounter{equation}{0}
\def\theequation{S.\arabic{equation}}

\section{Supplemental Material}

\section{Solvability of the Energy Equations}\label{sec:solvability}
Here we aim to demonstrate that in the case of encoding cities in $k=\log_2(n)$ qubits, there always exist $2^{2k}$ coefficients for each $E_{ij}$ in (7) satisfying the set of all $n^2=2^{2k}$ equations in Eq.\eqref{Eij_equations}.
To this end, we need to prove that all these equations are independent or the corresponding equation matrix is full-rank. Supposing that the city $v+1$ is encoded by the binary representation of the number $v = \sum\limits_{i=1}^k 2^i v_i$ denoted by $ \mathbf{v} = (\overline{v_1 ... v_k})_{_2}$, we can write the matrix form of these equations as below:

\begin{align}
	& \overbrace{
		\begin{bmatrix}
			1 & 0,\dots,0,0 & 0,\dots,0,0 & \dots & 0 & \dots & 0 & \dots & 0 \\
			1 & 0,\dots,0,0 & 0, \dots,0,1 & \dots & 0 & \dots & 0 & \dots & 0\\
			\vdots & \vdots & \vdots &  & \vdots &  & \vdots & & \vdots \\
			1 & a_1, \dots, a_k & b_1, \dots, b_k & \dots & & \dots &  & \dots & 0\\
			\vdots & \vdots & \vdots &  & \vdots &  & \vdots & & \vdots \\
			1 & 1,\dots,1,0 & 1,\dots,1,1 & \dots & 1 & \dots & 1 & \dots & 0\\
			1 & 1,\dots,1,1 & 1,\dots,1,1 & \dots & 1 & \dots & 1 & \dots & 1\\
	\end{bmatrix}}^{\text{\large{E}}^{(i,j)}}
	\overbrace{
		\begin{bmatrix}
			a_0\\
			\vdots\\
			a_{m_1}\\
			\vdots\\
			a_{m_1,m_2}\\
			\vdots\\
			a_{m_1,...,m_t}\\
			\vdots\\
			a_{1,2,...,2k}\\
	\end{bmatrix}}^{\Vec{\text{\large{$a$}}}}\nonumber \\
	= &
	\begin{bmatrix}
		E_{i,j}(\mathbf{0},\mathbf{0})\\
		E_{i,j}(\mathbf{0},\mathbf{1})\\
		\vdots\\
		E_{i,j}(\mathbf{a},\mathbf{b})\\
		\vdots\\
		E_{i,j}(\mathbf{n-2},\mathbf{n-1})\\
		E_{i,j}(\mathbf{n-1},\mathbf{n-1})\\
	\end{bmatrix}
	=
	\begin{bmatrix}
		P\\
		\delta_{i,j-1} W_{1,2}\\
		\vdots\\
		\delta_{i,j-1} W_{a+1,b+1}\\
		\vdots\\
		\delta_{i,j-1} W_{n-1,n}\\
		P
	\end{bmatrix}.
\end{align}

Here, each row of the matrix $E^{(i,j)}$ corresponds to a pair of cities (including cases where both cities are the same) e.g. $\mathbf{y}:=(\overline{a_1 \dots a_k,b_1 \dots b_k})_{_2}$ formed by concatenating the cities represented by $(\mathbf{a})_{_2}= (\overline{a_1 \dots a_k})_{_2}$ and $(\mathbf{b})_{_2}= (\overline{b_1 \dots b_k})_{_2}$,  while each column corresponds to the parameter $a_{m_1,m_2,\dots,m_t}$ where $M_t:=\{m_1, ..., m_t\}$ is a subset of the set $M_{2k}=\{1,2,\dots, 2k\}$. Thus the matrix $E^{(i,j)}$ consists of $n^2$ rows and $2^{2k}=2^{2\log_2(n)}=n^2$ columns.
According to Eq.\eqref{Eij}, the entry in the row $\mathbf{y}$ and the column $M_t$ is determined by multiplying the digits at the positions $m_1, ..., m_t$ of the binary string $\mathbf{y}$ i.e. $y_{m_1} \times ... \times y_{m_t}$.

We sort the columns in ascending order of their corresponding subset sizes, ranging from the empty set to the full set ($0\leq t \leq 2k$).
Then, we reorder the rows such that if the \(i\)-th column corresponds to the subset \(M_t\), then the \(i\)-th row corresponds to the bit string \(\mathbf{y}\), where \(y_{m_1}=\dots=y_{m_t}=1\) and the rest are 0.
By this arrangement of rows and columns, it is straightforward to verify that the diagonal entries are all 1, as \(y_{m_1}\times \dots \times y_{m_{t}}=1\). To compute the entries above the diagonal, we need to multiply the digits at the positions $M'_{t'\geq t} = \{m'_1, ..., m'_{t'}\}\neq M_t$ of \(\mathbf{y}\). Given that \(y_{m_1}=\dots=y_{m_t}=1\) and the rest are 0,  it follows that \(y_{m'_1}\times \dots \times y_{m'_{t'\geq t}}=0\). Consequently, the matrix takes the form of a lower triangular matrix, and its determinant equals the product of the diagonal entries, all of which are 1. Hence, the determinant is nonzero, and the matrix $E^{(i,j)}$ is a full-rank matrix.

\end{document}